\documentstyle[11pt,newpasp,twoside,epsf]{article}
\def\gtsima
{\hbox{\raise0.5ex\hbox{$>\lower1.06ex\hbox{$\kern-1.07em{\sim}$}$}}}
\def\ltsima
{\hbox{\raise0.5ex\hbox{$<\lower1.06ex\hbox{$\kern-1.07em{\sim}$}$}}}

\def\edcomment#1{\iffalse\marginpar{\raggedright\sl#1\/}\else\relax\fi}
\newcommand{\etal}{{et al.~}}
\reversemarginpar

\begin{document}
\title{The Nuclear Structure of the \\ Sagittarius Dwarf Spheroidal Galaxy}
\author{Lorenzo Monaco} \affil{Dip. di Astronomia, Universit\`a di
Bologna, Via Ranzani 1, 40127,  Bologna, ITALY \\ INAF - Osservatorio
Astronomico di Bologna, Via Ranzani 1, 40127, Bologna, 
ITALY} 

\author{Michele Bellazzini} \affil{INAF - Osservatorio Astronomico di
Bologna, Via Ranzani 1, 40127, Bologna,  ITALY}

\author{Francesco R. Ferraro} \affil{Dip. di Astronomia, Universit\`a
di Bologna, Via Ranzani 1, 40127,  Bologna, ITALY}

\author{Elena Pancino} \affil{INAF - Osservatorio Astronomico di
Bologna, Via Ranzani 1, 40127, Bologna, ITALY}

\begin{abstract}
We present a study of the central parts of the 
Sagittarius dwarf spheroidal galaxy (Sgr). We found a clear overdensity of 
Sgr's stars around M~54 (hereafter NS). 
NS is well represented by a King model and 
it has the characteristics of a typical dwarf elliptical nucleus. 
Whether this means that M~54 has spiraled into the potential well 
of NS or that M~54 is the real nucleus and NS has formed 
into its potential wells, remains an open question to be addressed.   
\end{abstract}

\section{Introduction}

Dwarf galaxies are considered the building blocks of the hierarchical merging 
process which is thought to be the fundamental mechanism for the formation of
large galaxies.
Among dwarf galaxies, dwarf ellipticals are the most common type of galaxy 
in the nearby universe (Ferguson \& Binggeli~1994, hereafter FB94). 
Hence, the comprehension of their structural and evolutionary characteristics 
is a major task of modern cosmology.

A characteristic of many dE is an enhancement of the surface brightness in a 
small central region, called {\it nucleus} (N). Such feature defines the
sub-class of nucleated dwarf elliptical (dE,N). 
The observed nuclei have
surface brightness profiles similar to globular clusters, they seem to share with
globulars the same general {\em surface brightness - absolute magnitude} relation
and their luminosity function overlaps the luminosity range covered by
globular clusters (FB94, Durrell 1997). 
Hence, the origin of dE,N
nuclei is generally reconduced to two possible mechanisms, both related with
massive star cluster: i.e. (a) the decay of the orbit of a pre-existing globular
toward the tip of the galactic potential well, driven by dynamical friction, or
(b) the {\em in situ} formation of a giant cluster from gas fallen to the 
center of the galaxy (see Durrell 1997 and references therein).
It has also been suggested that the capture hypothesis is more appealing to explane
faint nuclei, while the brightest ones may be better understood within the second
scenario. dE nuclei may be also related with the Ultracompact Dwarf Galaxies 
(UCD) recently discovered in the Fornax cluster (Drinkwater et al. 2003).
In summary, the phenomenon of dE nucleation is far from being well understood and
it is subject of continuous investigation (see, e.g. Stiavelli et al.(2001) and references therein). 
A local example would certainly provide a deeper
insight of the phenomenon, but the only acclaimed case in the Local Group is
M~32, a companion of the Andromeda galaxy whose stellar content may be studied in
some detail only with HST (Grillmair et al. 1996).

On the other hand, the Sagittarius dwarf spheroidal galaxy (Sgr) 
(Ibata \etal 1994)
is the most nearby Galactic satellite and since its discovery many authors
suggested and discussed the hypothesis that the associated massive globular
cluster M~54 is the nucleus of Sgr (Sarajedini \&
Layden 1995, Bassino \& Muzzio 1995, Layden \& Sarajedini 2000, hereafter LS00).
If the presently observed Sgr dSph is the relic of a previous dE,N it may provide
an excellent local testbed to study the nucleation process. Hence the detailed
study of its inner regions may have valuable spin-offs in this field. 

\section{The nuclear structure}
It is well known that the Sgr galaxy hosts a composite stellar population 
with stars spanning a pretty large range of ages (from $\ltsima$1 to 
$\gtsima$10 Gyr) and metallicities (from [Fe/H]$\ltsima$-1.5 to [Fe/H]$\simeq$0.0) 
(Monaco et al. 2002, hereafter Pap-I, LS00, Monaco et al. 2003, 
Bellazzini et al. 1999a,b).  
However its stellar content is dominated by a quite metal-rich population
($[M/H]\simeq -0.4/-0.6$) with an age of $\sim 4-6$ Gyr (LS00, Pap-I).
As a consequence, the color-magnitude diagram (CMD) of Sgr shows the typical 
features of an old metal-rich population with a clearly defined Red Clump of
He-burning stars and a cool Red Giant Branch (RGB). 
On the other hand M~54 is a quite old and metal poor globular cluster 
(LS00) with an extended blue Horizontal Branch and
a steep RGB. Therefore, the evolved stars of the two systems may be easily
discriminated on the CMD to study the respective spatial distributions. 

In Pap-I we presented a large photometric database containing position and photometry 
of $\sim$490,000 sources down to $V\sim 23$ in a $1 \times 1$ deg$^2$ field centered 
on M~54. From this sample we selected, on the basis of the CMD morphology, 
stars representative of M~54 and of the Sgr dominant population from the 
Pap-I database.  

Both selected samples 
are composed only of RGB stars and have the same limiting magnitude. Hence,
according to the evolutionary flux theory (Renzini \& Fusi Pecci 1988) they approximately trace
the same fraction of the total light of each system. 
For both samples we computed the star counts density profile. 
Since star counts are 
proportional to luminosity (Renzini \& Fusi Pecci 1988), we can convert the
projected 
star density (N/area) into the surface brightness: $\mu_{V}=-2.5\times Log(N/area)+ c$
where the value of the constant c is obtained by fitting the radial 
profile of M~54 with a King model 
(King 1962, using the structural parameters tabulated in Trager \etal 1995)
and normalizing the central surface brightness to the value reported in the literature, 
$\mu_{V,0}=14.75$~mag/arcsec$^2$.

The obtained surface brightness radial profiles are plotted on the left panel 
of Figure 1, as well as the model used to obtain the 
normalization. It can be appreciated that the model is a pretty good
representation of the profile of M~54 (assuming that the difference at r$<15\arcsec$ from the 
center of M~54 are due to incompleteness effect).
As can be seen, also the Sgr's sample shows a quite sharp central structure
(hereafter NS).   
\begin{figure}
\plottwo{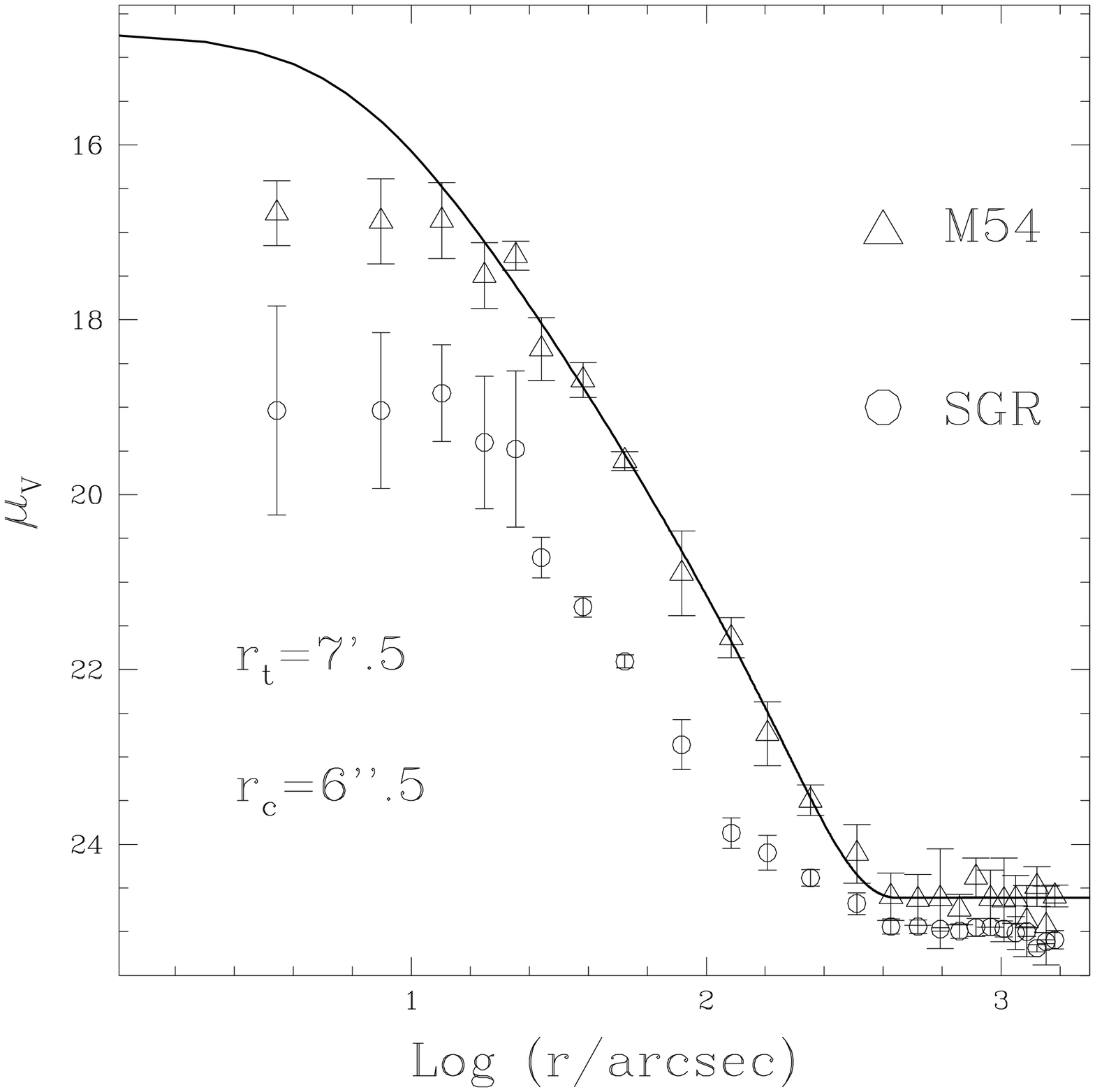}{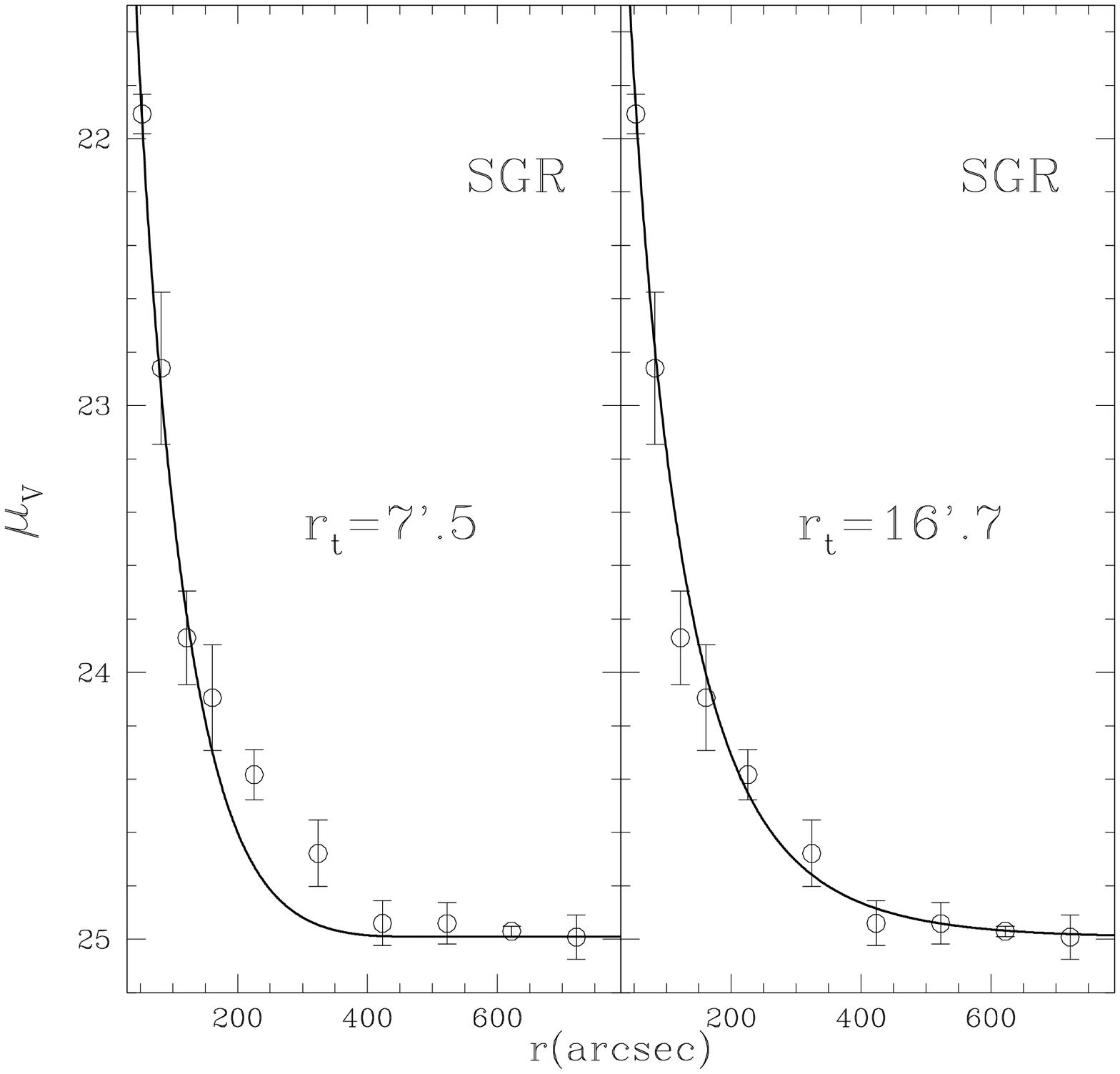}
\caption{Left hand panel: Radial surface
brightness profile obtained from a sample of M~54 (triangles) and Sgr (circles)
stars. The continuous line is the sum of a King model having the structural
parameters of M~54 (Trager \etal 1995) plus a constant component.
Right hand panels: The observed profile of Sgr is fitted by 
the sum of a King model plus a constant component. 
A good match is achieved in the right
panel where we used a tidal radius larger than the one of M~54.}
\end{figure}

We estimate the integrated magnitude of NS to be 
$-10.01\ltsima M_V\ltsima-8.2$, and the surface brightness of Sgr 
at the core radius of the main body to be $\mu_{V,c}\simeq26$~mag/arcsec$^2$ (using the model 
computed by Majewski et al. 2003).
These values place NS exactly in the locus occupied by dE,N in Figure 1 of 
Drinkwater \etal (2003) (Surface brightness of the envelope {\it vs} Core 
luminosity). 

Is M~54 or NS the nucleus of Sgr? Are M~54 and NS part of the same structure?
As can be seen in the right panels of Figure 1, a good match of the NS surface brightness
profile can only be obtained with a King model having a tidal radius larger 
than the one of M~54. A Kolmogorov-Smirnov test confirms that the two 
distributions are incompatible each other. We conclude that 
M~54 and NS do not share the same radial profile. Moreover 
M~54 is a metal poor Globular Cluster and its integrated colors are bluer 
than the Sgr field, while 
Galactic nuclei usually have colors similar to that of the 
surrounding galaxy field (FB94), even if a few nuclei bluer than 
the field do exist (Durrell 1997).
On the other hand, of course, NS has the same color of the surrounding galaxy. 
Moreover, if M~54 is the nucleus of Sgr, the difference in the central surface brightness of the nucleus with 
respect to the main body is $\Delta \mu_{V,0}^{Sgr-M~54}\simeq 10.25$~mag/arcsec$^2$, while
none of the 5 dE,N analysed by Geha et al.(2002)
and at most one of the 14 analysed by Stiavelli et al.(2001) show such a large difference. 
A lower $\Delta \mu_{V,0}^{Sgr-NS}$ value, more typical of dE,N, value is obtained for NS.  

It is thus safe to conclude that {\it NS shows the typical characteristics of a galactic
nucleus}, while the same statement does not apply to M~54.
 
While M~54 is generally considered the nucleus of Sgr, we now suggest the 
possibility that NS is the actual 
nucleus of Sgr and that M~54 has spiraled into the potential well of the 
nucleus. If this scenario (which needs further investigations) is correct,
galactic nuclei were 
finally demonstrated to be a distinct class of object with respect to Globular 
Clusters.

\end{document}